\documentclass[12pt,preprint]{aastex}

\shorttitle{Water formation} \shortauthors{Ioppolo et al.}

\begin{document}

\title{Laboratory evidence for efficient water formation in interstellar ices}

\author{S.~Ioppolo, H.~M.~Cuppen, C.~Romanzin, E.~F.~van Dishoeck and H.~Linnartz}
\affil{Raymond \& Beverly Sackler Laboratory for Astrophysics,
Leiden Observatory, Leiden University, PO Box 9513, 2300 RA Leiden,
              The Netherlands}

\begin{abstract}
Even though water is the main constituent in interstellar icy mantles, its chemical origin is not well understood. Three different formation routes have been proposed following hydrogenation of O, O$_2$, or O$_3$, but experimental evidence is largely lacking. We present a solid state astrochemical laboratory study in which one of these routes is tested. For this purpose O$_{2}$ ice is bombarded by H- or D-atoms under ultra high
vacuum conditions at astronomically relevant temperatures ranging
from 12 to 28 K. The use of reflection absorption infrared
spectroscopy (RAIRS) permits derivation of reaction rates and shows efficient formation of H$_{2}$O (D$_{2}$O) with
a rate that is surprisingly independent of temperature. This formation route  converts O$_2$ into H$_2$O via H$_2$O$_2$ and is found to be orders of magnitude more efficient than previously assumed. It should therefore be considered as an important channel for interstellar water ice formation as illustrated by astrochemical model calculations.
 \end{abstract}

   \maketitle

%________________________________________________________________

\section{Introduction}
Solid water ice has been observed on the surfaces of many different astronomical objects. In the Solar System it is found on planets and
minor bodies such as comets, trans-Neptunian objects and
Centaurs. In dense, cold interstellar clouds, infrared observations show that interstellar dust grains are covered with water-rich ices \citep[e.g.]{Gillett:1973,Gibb:2004,Pontoppidan:2004}. 
The formation of these ice mantles is especially important in the process of star and planet formation, when a large fraction of heavy elements can be depleted onto grains. In the dense cloud phase water layers form on the bare grain surfaces. Then during the gravitational (pre-)collapse, virtually all gas phase species freeze-out on top of these water layers resulting in a CO dominated layer that likely also contains traces of O$_2$ but very little water.

The observed H$_2$O ice abundance cannot be explained by direct accretion from the gas phase only. The exact mechanism by which water ice is formed is not understood.  The Herschel Space Observatory, to be launched in the near future, will provide important new information on gaseous water in interstellar space and will measure quantitatively the water abundance as a function
of temperature, UV field and other parameters. Furthermore, the Photodetector Array Camera and Spectrometer (PACS) on Herschel will cover the 62 $\mu$m band of solid H$_2$O. In this way Herschel will provide a unique opportunity to observe the bulk of the water bands that are unobservable from the ground and relate them to Spitzer and groundbased mid-IR observations of ices in protostellar envelopes and protoplanetary disks. Understanding the processes by which water forms and why it is not formed under other circumstances will be essential for the interpretation of these data.

\cite{Tielens:1982} proposed a reaction scheme in which water ice is formed on the surfaces of grains via three different routes: hydrogenation of O, O$_2$, and O$_3$. Models predict that water can indeed be formed through  such reactions in dense clouds \citep[e.g.]{Tielens:1982,dHendecourt:1985,Hasegawa:1993,Cuppen:2007}. Using a Monte Carlo approach \cite{Cuppen:2007} showed that the contributions of the different formation channels to water ice formation as well as its abundance strongly depend on the local environment. 
However, the initial reaction scheme with the corresponding rates as proposed by \cite{Tielens:1982} is based on old, in some cases outdated, gas phase data of the equivalent reactions. Progress has been severely hampered by the lack of realistic experimental simulations of these low-temperature, solid state reactions. Preliminary laboratory studies of water
synthesis  testing the first reaction channel have been reported by
\cite{Hiraoka:1998} and by \cite{Dulieu:2007}. Both groups
investigated the products of D- and O- reactions on an ice substrate
(N$_{2}$O and H$_{2}$O, respectively) using temperature programmed
desorption (TPD). In experiments exclusively using this technique, it
is hard to rule out any H$_{2}$O formation during warm up. Furthermore,
quantitative interpretation can be tricky because unstable species like H$_2$O$_2$ are destroyed in the mass spectrometer upon ionization leading to an artificially enhanced H$_2$O/H$_2$O$_2$ ratio.

The present work  focuses on the H + O$_2$ channel in which  O$_{2}$  is converted to H$_{2}$O via H$_2$O$_2$:
\begin{equation}
\rm{H} + \rm{O}_{2} \rightarrow \rm{HO}_{2}
\label{H+O2->HO2}
\end{equation}
\begin{equation}
\rm{H} + \rm{HO}_{2} \rightarrow
\rm{H}_{2}\rm{O}_{2}
\label{H+HO2->H2O2}
\end{equation}
\begin{equation}
\rm{H} + \rm{H}_{2}\rm{O}_{2} \rightarrow
\rm{H}_{2}\rm{O} + \rm{OH}
\label{H+H2O2->H2O+OH}
\end{equation}
\begin{equation}
\rm{H} + \rm{OH} \rightarrow \rm{H}_{2}\rm{O}
\label{H+OH->H2O}
\end{equation}
According to \cite{Cuppen:2007}  this channel is, together with the O$_3$ channel, responsible for water formation in cold, dense clouds. The exposure of O$_{2}$ ice to hydrogen and
deuterium atoms is investigated by means of reflection absorption infrared
spectroscopy (RAIRS) and TPD. These techniques allow one to determine formation yields and  the
corresponding reaction rates. The present work comprises a study of hydrogenation and deuteration reactions of O$_{2}$ ice for
different temperatures between 12 and 28 K, i.~e.~, up to the desorption temperature of O$_{2}$ \citep{Acharyya:2007}. The formation of H$_2$O and H$_2$O$_2$ is observed at all temperatures. An optimum yield is found at 28 K.

\section{Experimental}
Experiments are performed using an ultra-high vacuum set-up ($P  < 5
\times 10^{-10}$ mbar) which comprises a main chamber and an atomic
line unit. The set-up is discussed in more detail in \cite{Fuchs:prep}. The main chamber contains a gold coated copper substrate
($2.5 \times 2.5$ cm$^{2}$) that is in thermal contact with the
cold finger of a 12 K He cryostat. The temperature can be varied
with 0.5 K precision between 12 and 300 K. A precision leak valve is
used to deposit O$_{2}$ ($99.999\%$ purity, Praxair) on the substrate. Ices are grown at
$45^{\circ}$ with a flow of $1 \times 10^{-7}$ mbar s$^{-1}$ where
$1.3 \times 10^{-6}$ mbar s$^{-1}$ corresponds to 1 Langmuir (L)
s$^{-1}$. In order to compare results from different experiments,
the thickness of the O$_{2}$ ice is 75 L  for all samples studied 
and the substrate temperature is kept
at 15 K during the deposition. An O$_2$ ice of 75 L consists of roughly 30 monolayers. This thickness is chosen to exclude substrate induced effects. Because a diatomic homonuclear molecule like O$_{2}$ is infrared
in-active, gas phase O$_{2}$ is monitored during the deposition by a
quadrupole mass spectrometer (QMS). After deposition at 15 K the ice is slowly cooled down or
heated (1 K min$^{-1}$) until a selected temperature is reached. Systematic
studies are performed for different temperatures between 12 and 28
K.

H(D)-atoms are produced in a well characterized thermal-cracking
device \citep{Tschersich:1998,Tschersich:2000}. A second precision
leak valve is used to admit H$_{2}$ (D$_{2}$) molecules ($99.8\%$
purity, Praxair) into the gas cracking line. In each
experiment the H + H$_{2}$ (D + D$_{2}$) flow through the capillary
in the atomic line is $ 1 \times 10^{-5}$ mbar s$^{-1}$ and the
temperature of the heated tungsten filament, which surrounds the gas
cracking pipe, is about 2300 K. The dissociation rate and the atomic
flux depend on the pressure and temperature 
\citep{Tschersich:2000} and are kept constant during all the
experiments. A nose-shaped quartz pipe is placed along the path of the atomic
beam in order to cool down H(D)-atoms to room temperature before
reaching the ice sample by collisions \citep{Walraven:1982}.
The 
H(D)-atomic flux nearby the sample is estimated, within 50 \%, as $ 5 \times 10^{13}$ cm$^{-2}$ s$^{-1}$. At temperatures of 12 K and
higher, no blocking of surface processes by the presence of H$_{2}$ is expected in the ice.

The newly formed species after hydrogenation (deuteration) of O$_{2}$ ice
are monitored by RAIRS
using a Fourier transform infrared spectrometer (FTIR) running at a
spectral resolution of 4 cm$^{-1}$ in the range between 4000
and 700 cm$^{-1}$ ($2.5 - 14$ $\mu$m). Typically the ice is exposed to the H (or D) beam for 3 (or 2) hours and IR spectra are acquired every few minutes. 

Systematic control experiments have been performed in order \emph{(i)} to
unambiguously confirm that the products are formed by surface processes and not by gas phase reactions, \emph{(ii)} to check that any water present in the system does not affect the final results and \emph{(iii)} to verify that water and H$_2$O$_2$ formation
occurs in the solid phase after H(D)-atom bombardment and not by
H$_{2}$(D$_{2}$)-molecules addition. For \emph{(i)} co-deposition experiments are undertaken in which H and O$_2$ are deposited simultaneously. Water is only formed if the surface temperature is below the desorption temperature of oxygen, confirming that the presence of the oxygen ice is required for this reaction sequence to occur. Point \emph{(ii)} is verified by using inert initial substrates like N$_2$ ice to estimate the background water contribution, as well as by using different  isotopologues ($^{18}$O$_2$, $^{15}$N$_2$ and D). Finally \emph{(iii)} is checked by using pure H$_{2}$(D$_{2}$)-beams, i.~e.~without any H(D) present.

\section{Results}
\begin{figure}
\plottwo{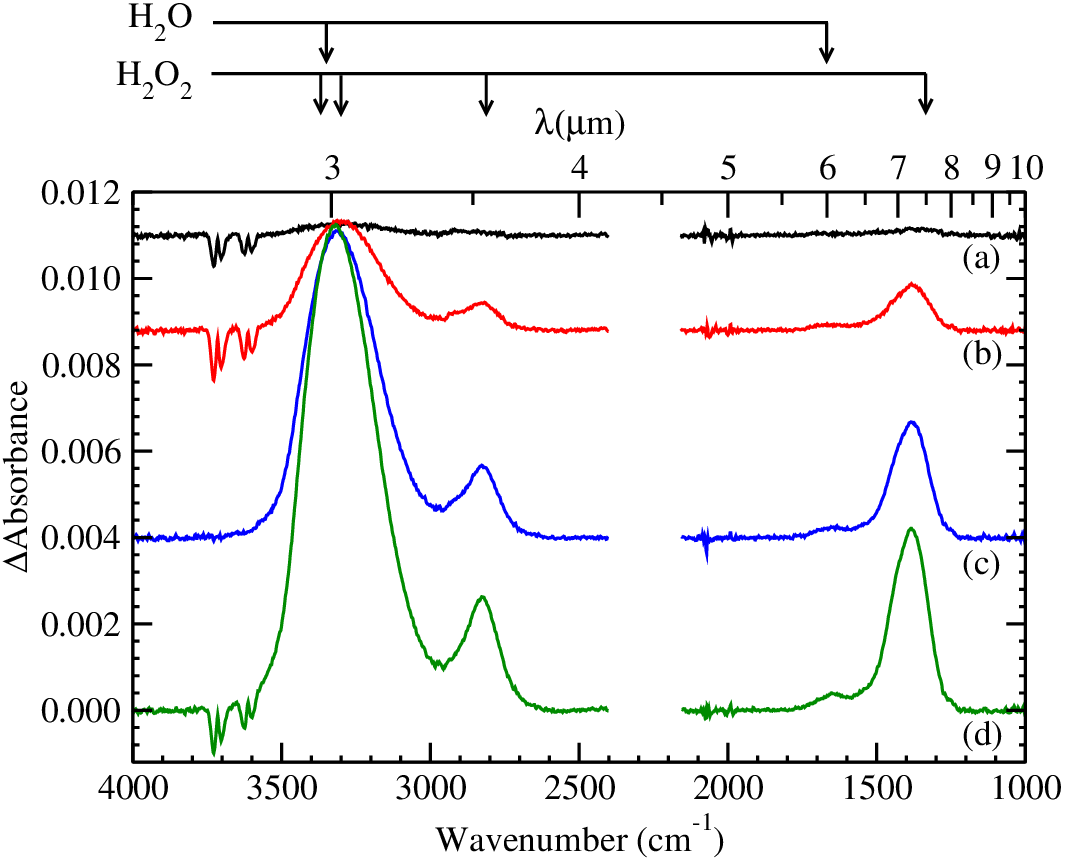}{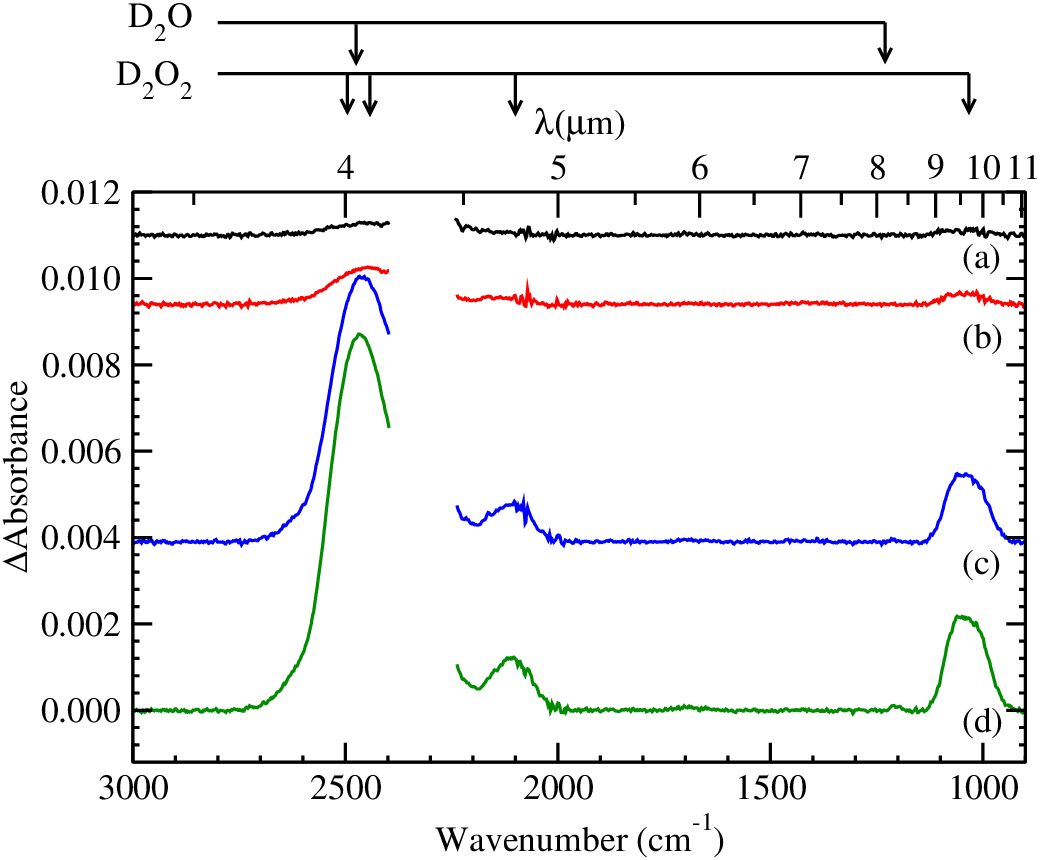}
\caption{RAIR spectral changes of the O$_2$ ice as a function of  H-atom (left) and D-atom (right) bombardment at 25 K. Spectra at a H(D)-atom fluence of (a) $4\times 10^{15}$, (b) $4\times 10^{16}$, (c) $1\times 10^{17}$, and (d) $2\times 10^{17}$ cm$^{-2}$ are given. }
\label{spectra}
\end{figure}

The formation of both H$_2$O$_2$ and H$_2$O ice is confirmed by the appearance of their infrared solid state spectral signatures. Figure \ref{spectra} shows  typical RAIRS results for hydrogenation and deuteration of O$_2$ ice at 25 K. From top to bottom a time sequence of four spectra is plotted. These spectra are difference spectra with respect to the inital oxygen ice. However, since our initial oxygen ice only consists of 30 ML, no features due to the intrisically very weak O$_2$ feature \citep{Ehrenfreund:1992, Bennett:2005} are observed in the original spectrum. Both the H$_2$O and  H$_2$O$_2$ clearly grow in time as the H-fluence (H-flux $\times$ time) increases. Similar features appear for the deuteration experiment, although here clearly less D$_2$O is formed.
After fitting the infrared spectra with a straight baseline the
column density (molecules cm$^{-2}$) of the newly formed
species in the ice is calculated from the integrated intensity
of the infrared bands using a modified Lambert-Beers equation
\citep{Bennett:2004}.
In the range of our spectrometer, water ice has two candidate bands for determining its column density, at 3430
cm$^{-1}$  and  1650 cm$^{-1}$ (3 and 6 $\mu$m, respectively). Since
the strong 3430 cm$^{-1}$ feature overlaps with the 3250 cm$^{-1}$
band of H$_2$O$_2$, the weak feature at 1650 cm$^{-1}$ was chosen to
quantify the water column density. Since literature values of transmission
band strengths cannot be used in reflection measurements, an
apparent absorption strength
is obtained from a calibration experiment in which a pure water ice layer desorbs at constant temperature until the
sub-monolayer regime \citep{Oberg:2007}. The uncertainty in the band strengths remains within a factor of two.  Quantification of H$_2$O$_2$ is done using the 1350
cm$^{-1}$ band. As it is experimentally very hard to deposit  pure 
H$_2$O$_2$ ice, the apparent absorption strength has
to be obtained indirectly by assuming the ratio of the integrated
band strengths between the two bands in transmittance to be the same as in reflectance H$_2$O/H$_2$O$_2$ = 0.57
\citep{Gerakines:1995, Loeffler:2006}. The band
modes of solid D$_{2}$O and D$_{2}$O$_{2}$ ices have systematic
peak position shifts of $\sim$400 cm$^{-1}$ with respect to the H$_{2}$O and H$_{2}$O$_{2}$ bands.
The column density for deuterated species is obtained in a similar
way from a calibration experiment, while the absorption strength for D$_{2}$O$_{2}$ is estimated assuming that H$_2$O/H$_2$O$_2$ = D$_2$O/D$_2$O$_2$.

In Fig.~\ref{NH2O} the column densities of water and 
H$_2$O$_2$ are plotted as a function of the H-fluence (atoms
cm$^{-2}$) for different substrate temperatures. Figure \ref{ND2O} gives the equivalent for the deuterated species. The results for H$_2$O$_2$ and D$_2$O$_2$ are found to be very reproducible (errors within the symbols). Due to their low column densities the relative errors for H$_2$O and especially D$_2$O are larger. The H$_2$O$_2$ and D$_2$O$_2$ results all show the same initial linear increase followed by a very sharp transition to a steady state column density, with the steady state value increasing with temperature. 
The water results show similar behavior although the transition is not as sharp and the temperature dependence of the steady-state value is not as clear. The observation that the results for all temperatures show the same initial slope means that the rate of the reaction to H$_2$O$_2$ is temperature independent. The final yield is however temperature dependent and this depends on the penetration depth of the hydrogen atoms into the O$_2$ ices;  at higher temperatures H-atoms can penetrate deeper. This is discussed in more detail later.

During the preparation of this manuscript we received a preprint by
\cite{Miyauchi:2008} who performed a similar experiment for one single temperature (10 K). Our results for 12 K
turn out to be close to their results, apart from an absolute scaling due to different assumptions on the band strengths.

\section{Determining the reaction rates}
Reaction rates are obtained by fitting a set of differential equations to the time evolution curves of H$_{2}$O (D$_{2}$O) and H$_{2}$O$_{2}$ (D$_{2}$O$_{2}$). 
Usually a diffusive mechanism is considered to construct these equations. For H$_2$O$_2$ that is formed from oxygen and converted into water as given in Eqs.~\ref{H+O2->HO2}-\ref{H+OH->H2O} the rate equation would be 
\begin{eqnarray}
\frac{\rm{d}n_{\rm{H}_2\rm{O}_2}(t)}{\rm{d}t} &=&  (k_{\rm hop}^{\rm H} + k_{\rm hop}^{\rm O_2}) k_{1} n_{\rm H}(t) n_{\rm O_2}(t) \nonumber\\
&-&  (k_{\rm hop}^{\rm H} + k_{\rm hop}^{\rm H_2O_2}) k_{3} n_{\rm H}(t) n_{\rm H_2O_2}(t)
\end{eqnarray}
with $k_{\rm hop}^{\rm X}$ and $n_{\rm X}$ the hopping rate and column density of species X and $k_{1}$ and $k_{3}$ the rate constants of reactions 1 and 3. 

\cite{Awad:2005} and \cite{Miyauchi:2008} applied similar expressions to determine the rates of the formation of methanol and water, respectively. They solved these equations under the assumption that the atomic hydrogen abundance on the surface remains constant. In this way expressions for the surface abundances of the reaction products were obtained with only the effective reaction rates as fitting parameter
\begin{equation}
n_{\rm H_2O_2} = n_{\rm O_2}(0) \frac{\beta_{1}}{\beta_{1}-\beta_{3}} \left( \exp(-\beta_{3}t) - \exp(-\beta_{1}t)\right),
\end{equation}
with
\begin{equation}
\beta_{\rm 1} = (k_{\rm hop}^{\rm H} + k_{\rm hop}^{\rm O_2}) k_{1} n_{\rm H}.
\label{beta}
\end{equation}

\begin{figure}
\epsscale{0.45}
\plotone{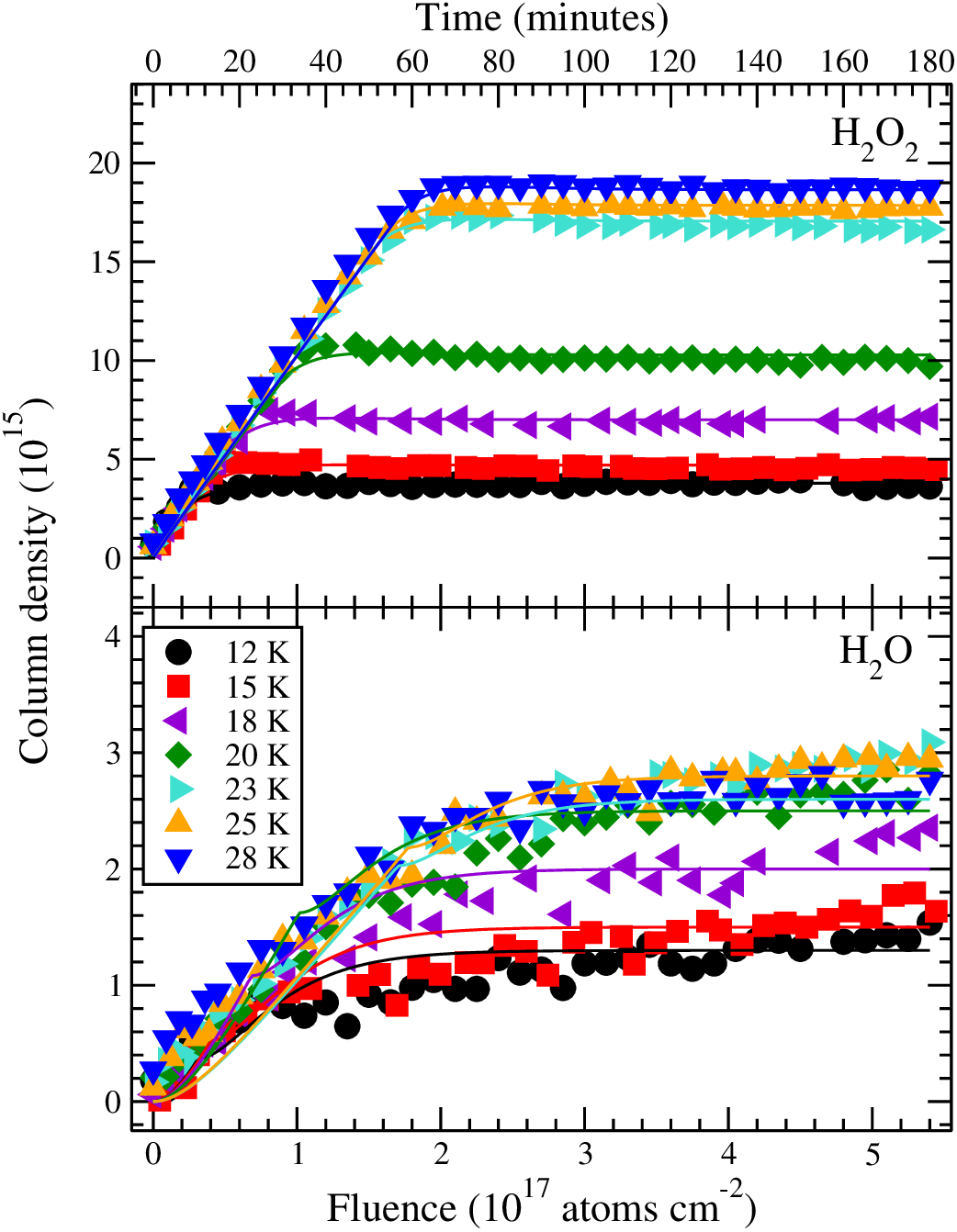}
\caption{The  H$_2$O$_2$ (top) and H$_2$O (bottom) column densities as a function of time and H-atom fluence for different O$_2$ ice temperatures. The symbols indicate experimental data, the solid lines represent the fitted model.}
\label{NH2O}
\end{figure}

\begin{figure}
\epsscale{0.45}
\plotone{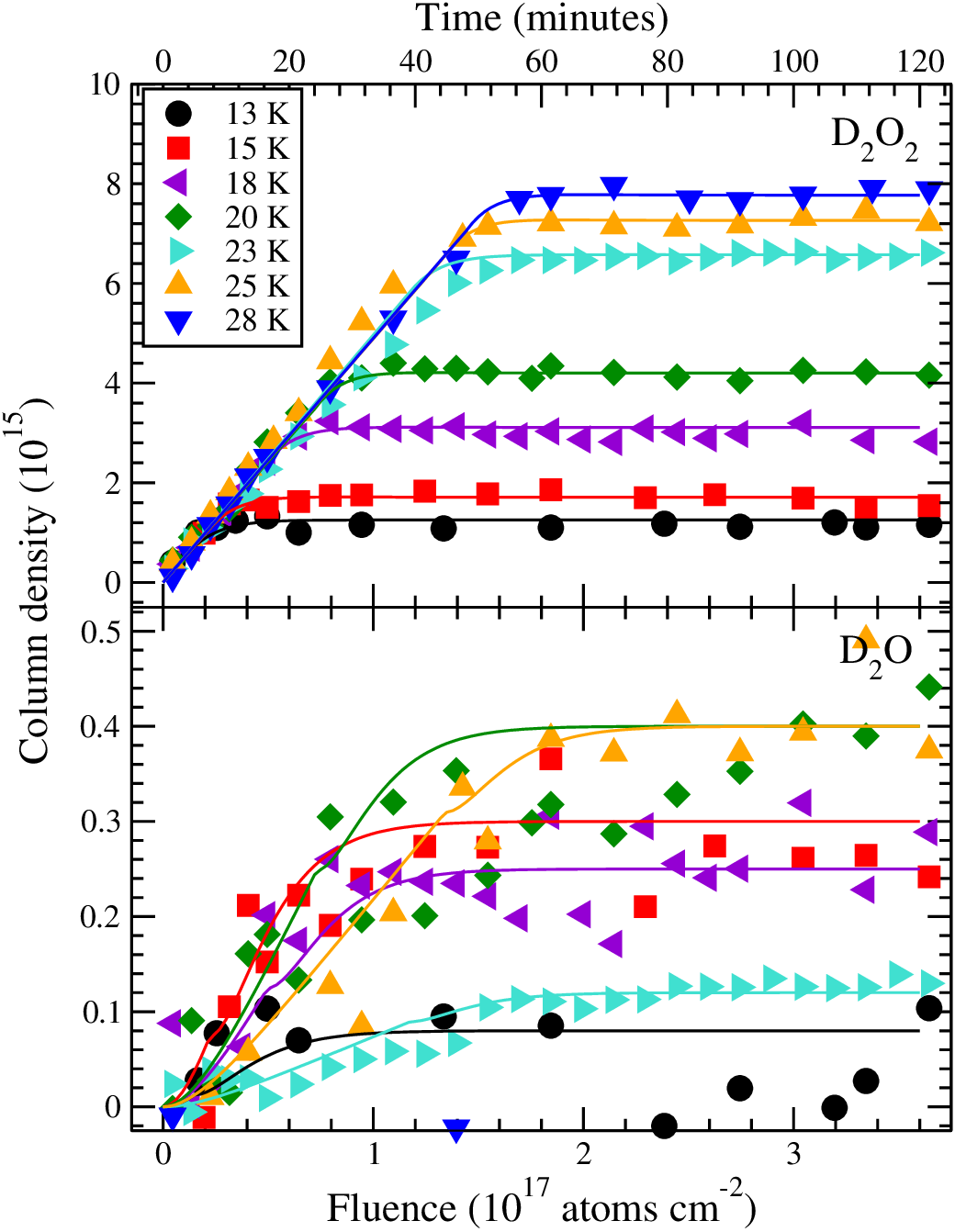}
\caption{Similar to Fig.~\ref{NH2O} for D$_2$O$_2$ (top) and D$_2$O (bottom).}
\label{ND2O}
\end{figure}

Fitting this expression to the experimental data points in Figs.~\ref{NH2O} and \ref{ND2O} gives a very poor agreement because the model has an exponential behavior whereas the experiments for several temperature values clearly do not. This would moreover result in a different rate for H + O$_2$ for each temperature whereas the experimental curves  show that the rate is independent of  temperature. 
For this reason we decided to use a different model. We consider two regimes. In the first regime ($ t < t_t$) the hydrogen atoms get trapped into the ice with an efficiency that is independent of  temperature. Once an atom is trapped it can diffuse efficiently and find an oxygen molecule to react with. This results in a zeroth order rate. Once nearly all oxygen molecules within the penetration depth are converted to H$_2$O$_2$ ($ t > t_t$), diffusion becomes rate limiting  and a diffusive mechanism for the reaction to H$_2$O$_2$ and H$_2$O is applied. Since now the ice is changed from an O$_2$ ice to an H$_2$O$_2$ ice the penetration depth of the H-atoms into the ice will change as well.
This model is described by the following set of equations
\begin{eqnarray}
n_{\rm{H}_2\rm{O}_2} & = & \beta_{1}t \frac{p_{\rm{O}_2}-p_{\rm{H}_2\rm{O}_2}}{p_{\rm{O}_2}} + \frac{\beta_{1}}{\beta_3}\frac{p_{\rm{H}_2\rm{O}_2}}{p_{\rm{O}_2}}\left(1-\exp(-\beta_3 t)\right)\\
n_{\rm{H}_2\rm{O}} & = & \beta_1'\frac{p_{\rm{H}_2\rm{O}_2}}{p_{\rm{O}_2}}\left(t-\frac{1-\exp(-\beta_3t)}{\beta_3}\right)
\end{eqnarray}
for $ t < t_t$ and 
\begin{eqnarray}
n_{\rm{H}_2\rm{O}_2} & = & p_{\rm{H}_2\rm{O}_2}\frac{\beta_{1}'}{\beta_{1}'-\beta_3}[\exp(-\beta_3(t-t_t))-\exp(-\beta_1(t-t_t))] \nonumber \\
& + &(p_{\rm{O}_2}-p_{\rm{H}_2\rm{O}_2}-n_{\rm{H}_2\rm{O}_2}(t_t))(1-\exp(-\beta_1(t-t_t)))\nonumber \\
&+& n_{\rm{H}_2\rm{O}_2}(t_t)\\
n_{\rm{H}_2\rm{O}} & = &  n_{\rm{H}_2\rm{O}}(t_t) + (p_{\rm{H}_2\rm{O}_2}-n_{\rm{H}_2\rm{O}}(t_t)) \times \nonumber\\
&&\left(1+\frac{\beta_3\exp(-\beta_1(t-t_t)) - \beta_1\exp(-\beta_3(t-t_t))}{\beta_1-\beta_3}\right) 
\end{eqnarray}
for $ t > t_t$
 with $p_{\rm{X}}$ the penetration depth of H-atoms into ice X in units of column density, $\beta_1$ and $\beta_1'$ the effective rates of reaction 1 and $\beta_3$ the rate of reaction 3. $\beta_1$ and $\beta_3$ represent effective diffusive rates that include both the diffusion rate and the reaction rate, whereas $\beta_1'$ is mainly determined by the hydrogen flux times the efficiency of H trapping into the ice. 

Fitting this model to the data, the three rates are found to be independent of temperature. We therefore apply the same average rate to describe the results at all temperatures and only the two penetration depths are allowed to vary between the experiments. The resulting curves are indicated by the solid lines in Figs.~\ref{NH2O} and \ref{ND2O}. The obtained rates are given in Table \ref{para}. The penetration depths are the steady state values in Figs.~\ref{NH2O} and \ref{ND2O}. These clearly increase with temperature to very high values. This suggests that as the O$_2$ ice reaches its desorption temperature ($\sim$30 K) the structure becomes more open and the O$_2$ molecules more mobile, allowing the hydrogen atoms to penetrate deeply into the ice. The H$_2$O$_2$ structure on the contrary is much more dense and rigid and consequently the H/D-atoms cannot penetrate more than a few monolayers even at the highest temperatures. The temperature effect is also much less prominent in cases where the ice is comfortably below its desorption temperature. 

A clear difference in penetration depth between the hydrogenation and deuteration experiments can be observed. For the oxygen penetration depth a difference of a factor of two is found whereas for hydrogen and deuterium peroxide the difference can be even as high as a factor of six.

Like in the models by \cite{Awad:2005} and \cite{Miyauchi:2008}, the rates $\beta_1$ and $\beta_3$ are the products between the rates of the reactions and the hydrogen surface abundance. The latter is assumed to be constant during the experiment and is the overall result of accretion, desorption and
reaction with both H and O$_2$. A comparison of $\beta_1'$ and $\beta_1$ indicates that the reaction with H only plays a minor role.
 
The uncertainties in $\beta_1$ and $\beta_3$ are mainly determined by the fit and are within 50\%. The error in  $\beta_1'$ due to the fit is much less, but here the main uncertainty is determined by the layer thickness. The error in the calibration of the layer thickness is a factor of two. 
Assuming that every hydrogen atom that gets trapped reacts with O$_2$  and considering that two hydrogen atoms are needed to convert O$_2$ to H$_2$O$_2$, the trapping efficiency can be determined from $\beta_1'$ and the flux ($ 5 \times 10^{13}$ cm$^{-2}$ s$^{-1}$). This  results in a trapping efficiency of $\sim$10\% for deuterium and $\sim $20\% for atomic hydrogen. This is very close to the sticking efficiency of hydrogen atoms to a water ice surface of $\sim$30\% under these circumstances \citep{Al-Halabi:2006}. The high efficiency of reaction 1 is consistent with recent theoretical studies of this reaction in the gas phase. \cite{Xie:2007} and \cite{Xu:2005} found that this reaction can proceed barrierless for certain incoming angles.
The rates $\beta_1$ and $\beta_3$ have very similar values. This suggests that also reaction 3 is very efficient and H$_2$O formation is only limited by the penetration depth of H$_2$O$_2$. 

The rate of reaction \ref{H+H2O2->H2O+OH} shows no significant isotope effect. This is in contrast with the results by \cite{Miyauchi:2008} who found a significant isotope effect for this reaction using the diffusive model to fit their 10 K results. If a large barrier is involved in reaction \ref{H+H2O2->H2O+OH}, barrier crossing would proceed via tunneling and an isotope effect is expected (Watanabe, private communications). The fact that we do not observe a (large) effect at 12-28 K either suggests that the barrier for this reaction is low or that other mechanisms for the formation of water should be taken into account in the model to fit the data. In a future paper we plan to address this question in more detail. For now the conclusion remains that the formation of water from O$_2$ and H is very efficient.

\section{Astrophysical discussion and conclusion}
The hydrogenation and deuteration experiments presented in this letter show an efficient mechanism to convert O$_2$ ice to H$_2$O$_2$ and ultimately H$_2$O. In the model that describes these experiments, the rate limiting steps for formation are the trapping of the hydrogen into the O$_2$ ice and the penetration depth of the hydrogen into the H$_2$O$_2$. Astrochemical models take reaction, diffusion and desorption barriers as input. Our data shows that the formation of at least H$_2$O$_2$ proceeds via a reaction with a barrier that is much lower than the value of 1200 K previously assumed by \cite{Tielens:1982}. It should therefore be  considered as a route for water formation on interstellar grain surfaces. 

Care should however be taken when extending these findings from a laboratory environment directly to interstellar ices. 
The temperature independence of the reaction rate that is observed, for instance, is directly due to the unique property of the O$_2$ ice that allows hydrogen to penetrate its structure and thereby preventing desorption back into the gas phase. 
In interstellar clouds the mantle surfaces would not consist of pure O$_2$ ice but of a mixture of different species with water as its main constituent  \citep{Whittet:1998}. The structure of these ``dirty'' ices and the binding energies to it would govern the desorption and diffusion behavior of adsorbates like O$_2$ and H. 
Experiments on the binding of atomic hydrogen on water ice surfaces showed that the surface abundance is generally dependent on temperature \citep{Perets:2005, Hornekaer:2003, Dulieu:2005}. Also hydrogenation experiments of CO showed that the decreasing H coverage for increasing temperature results in lower effective rates \citep{Watanabe:2003,Watanabe:2006A,Linnartz:2006,Cuppen:HK}. 
Here we discuss the implications of the present work to two different interstellar environments: hydrogenation of an apolar (water-poor) ice mantle after freeze-out and cold cloud conditions where ice is formed from direct deposition of H and O.

Observations show that interstellar ice mantles consist of polar (water-rich) and apolar (water-poor) layers \cite{Tielens:1991}.
Apolar ices are thought to form during freeze-out in the densest parts of the cloud. In dense gas, most of the atomic oxygen has converted to O$_2$ which can become subsequently a constituent of this polar phase. The lack of observed H$_2$O$_2$ and H$_2$O means that most of the O$_2$ in the ice does not react to produce H$_2$O$_2$ or H$_2$O. This laboratory work however shows that hydrogen atoms can penetrate deeply into O$_2$ ices and then react. In apolar interstellar ices the penetration depth is not determined by oxygen ice but by the main constituent of the ice mantle, CO. We therefore have performed additional laboratory experiments in which a mixture of CO and O$_2$ ice is exposed to H-atoms. These show indeed that only the top few layers are hydrogenated and that the main part of the ice stays intact in agreement with the observations. Details of these experiments are to be published in a future paper.

\begin{figure}
\epsscale{0.45}
\plotone{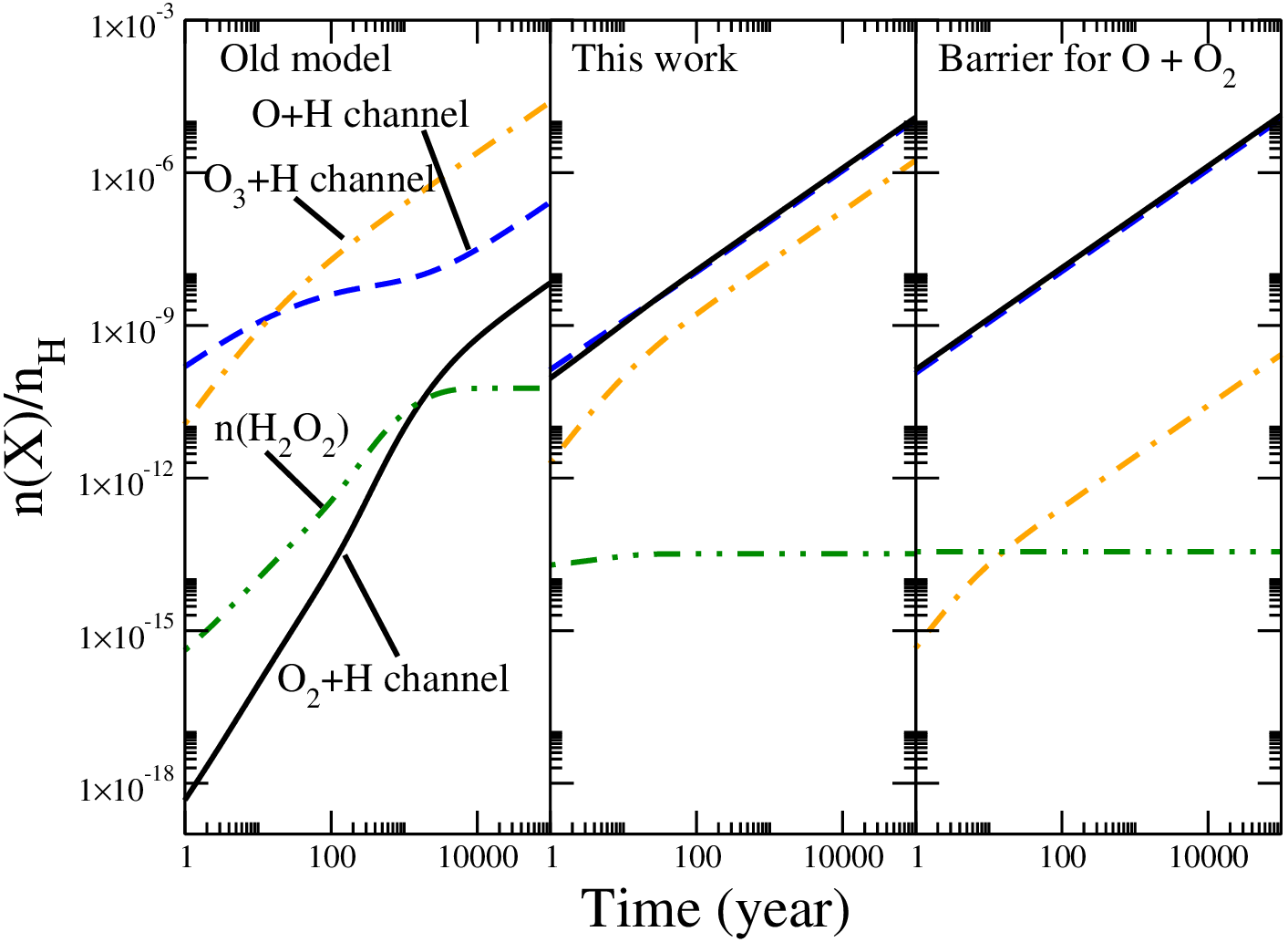}
\caption{The contributions of the three different channels to H$_2$O formation in three models and the H$_2$O$_2$ abundance, $n$(H$_2$O$_2$) from the O$_2$ + H channel. (left) Old network by \cite{Tielens:1982}, (middle) new network without barriers for reactions 1 and 3 and (right) new network with barrier for O + O$_2$ reaction. See text for details.}
\label{model}
\end{figure}

In cold and translucent clouds H- and O-atoms deposit onto the grain simultaneously. The species can then react  immediately and O$_2$ is converted all the way to water. Here the penetration depth observed in the laboratory becomes unimportant. What can we learn from the present experiments then? The fast reaction kinetics  justify treating H + O$_2$ and H + HO$_2$ in a similar way as H + H are treated in astrochemical models, with the difference that probably only hydrogen is mobile for reaction. The results further suggest that the continuing reactions leading to water proceed with high efficiency. 
The original grain surface network by \citep{Tielens:1982} includes two more water formation routes: via OH and via ozone. Under dense cloud conditions the ozone route was proposed to be the most efficient as is shown in the left panel of Fig.~\ref{model}. This figure plots the contributions for the three different H$_2$O formation channels using a reaction network limited to the three water formation routes and without any dissociation reactions. The model parameters are taken from model M1 by \cite{Ruffle:2000} and the initial conditions from \cite{Tielens:1982} are used ($n({\rm H}_2) = 5 \times 10^4$, $n({\rm H}) = 1.41$ and $n({\rm O}) = 2.36$ cm$^{-2}$). These calculations include a barrier for reactions 1 and 3. The present work shows that these barriers are negligible and the middle panel plots the same model calculation without these barriers. The figure clearly shows that the O$_2$ channel has a major contribution to the overall water formation rate.   Recent laboratory experiments by \cite{Sivaraman:2007} on the temperature-dependent formation of ozone by irradiation of oxygen ices by high energy electrons suggest that mobile oxygen atoms prefer the O + O pathway over O + O$_2$ even though the O/O$_2$ ratio in the ice is very small. This implies the presence of a barrier for the formation of ozone. The right panel gives the model results when a small barrier of 500 K for this reaction is considered. The contribution of the ozone channel is now further reduced.

Figure \ref{model} further shows the H$_2$O$_2$ abundance, $n$(H$_2$O$_2$), for all models as produced through the O$_2$ + H route. \cite{Boudin:1998} set an observational constraint of the  H$_2$O$_2$ ice abundance toward \object{NGC7538:IRS9} of 5.2\% with respect to the H$_2$O ice abundance. All models are consistent with this number.

\section{Acknowledgments}
Funding was provided by NOVA, the Netherlands Research School for
Astronomy and by a Spinoza grant and a VENI grant both from the Netherlands Organization
for Scientific Research, NWO. We thank Lou Allamandola, Xander Tielens and Stefan Andersson for
many stimulating discussions.

% for the bibliography, at the end

\newpage

\newpage

\newpage

\newpage

\begin{deluxetable}{lccccc}
\tabletypesize{\scriptsize}
\tablecaption{The reaction rates obtained by fitting the model. For the corresponding uncertainties see the text.\label{para}}
\tablewidth{0pt}
\tablehead{
 \colhead{}  & \colhead{$\beta_1'$} & \colhead{$\beta_1$} & \colhead{$\beta_3$}\\
\colhead{}   &   \colhead{[molec cm$^{-2}$s$^{-1}$]} & \colhead{[s$^{-1}$]} & \colhead{[s$^{-1}$] }
}
\startdata
H + O$_2$ & $5.4\times 10^{12}$ & $1.3 \times 10^{-3}$ & $1.8 \times 10^{-3}$ \\
D + O$_2$ & $2.5\times 10^{12}$ & $3.3 \times 10^{-3}$ & $2.7 \times 10^{-3}$ \\
\enddata
%% Text for table notes should follow after the \enddata but before
%% the \end{deluxetable}. Make sure there is at least one \tablenotemark
%% in the table for each \tablenotetext.
\end{deluxetable}

\end{document}